\documentclass{nic-series}
\usepackage{macros}
\usepackage{color}
\usepackage{wrapfig}

\begin{document} 

\thispagestyle{empty}

\vspace{-1.0cm}
\begin{flushright}
\large DESY 02--030
\end{flushright}
\vspace{1.0cm}

\begin{center}
{\huge Fundamental parameters of QCD}\\[12mm]
\vskip 0.5 cm
\vbox{
\centerline{
\epsfxsize=2.5 true cm
\epsfbox{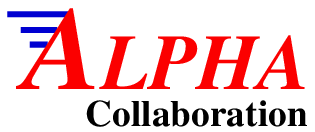}}
}
\vskip 0.5 cm

{\large Rainer Sommer$^1$ and Hartmut~Wittig$^2$}\\ [0.3cm]
$^1$DESY, Platanenallee 6, 15738 Zeuthen, Germany\\ [0.3cm]
$^2$DESY, Notkestrasse 85, 22603 Hamburg, Germany\\ [0.3cm]

\end{center}
\vspace{15mm}
\centerline{\bf Abstract}
\par\noindent{\small 
The theory of strong interactions, QCD, is described in terms of a few
parameters, namely the strong coupling constant $\alpha_s$ and the
quark masses. We show how these parameters can be determined reliably
using computer simulations of QCD on a space-time lattice, and by
employing a finite-size scaling method, which allows to trace the
energy dependence of $\alpha_s$ and quark masses over several orders
of magnitude. We also discuss methods designed to reduce the effects
of finite lattice spacing and address the issue of computer resources
required.}
\vspace{5cm}
\begin{center}
{\large (Contribution to {\it NIC Symposium 2001 -- Proceedings})}
\end{center}

\clearpage

\setcounter{page}{1}

\title{Fundamental parameters of QCD}

\author{
ALPHA Collaboration,   
Rainer Sommer \inst{1}  \and Hartmut Wittig \inst{2}}

\institute{DESY,\\
         Platanenallee 6, 15738 Zeuthen, Germany\\
         \email{rainer.sommer@desy.de}
         \and
         DESY,\\
         Notkestrasse 85, 22603 Hamburg, Germany \\
         \email{hartmut.wittig@desy.de} }

\maketitle

\def\alphastrong{\alpha_{\rm QCD}}
\newcommand{\red}{\color{red}}
\newcommand{\yel}{\color{yellow}}
\newcommand{\blu}{\color{blue}}
\newcommand{\gre}{\color{green}}
\newcommand{\bla}{\color{black}}
\newcommand{\mgt}{\color{magenta}}

\begin{abstracts}
The theory of strong interactions, QCD, is described in terms of a few
parameters, namely the strong coupling constant $\alpha_s$ and the
quark masses. We show how these parameters can be determined reliably
using computer simulations of QCD on a space-time lattice, and by
employing a finite-size scaling method, which allows to trace the
energy dependence of $\alpha_s$ and quark masses over several orders
of magnitude. We also discuss methods designed to reduce the effects
of finite lattice spacing and address the issue of computer resources
required.
\end{abstracts}
\section{The Standard Model of particle physics}

\begin{figure}
\begin{center}
\includegraphics[width=9.5cm]{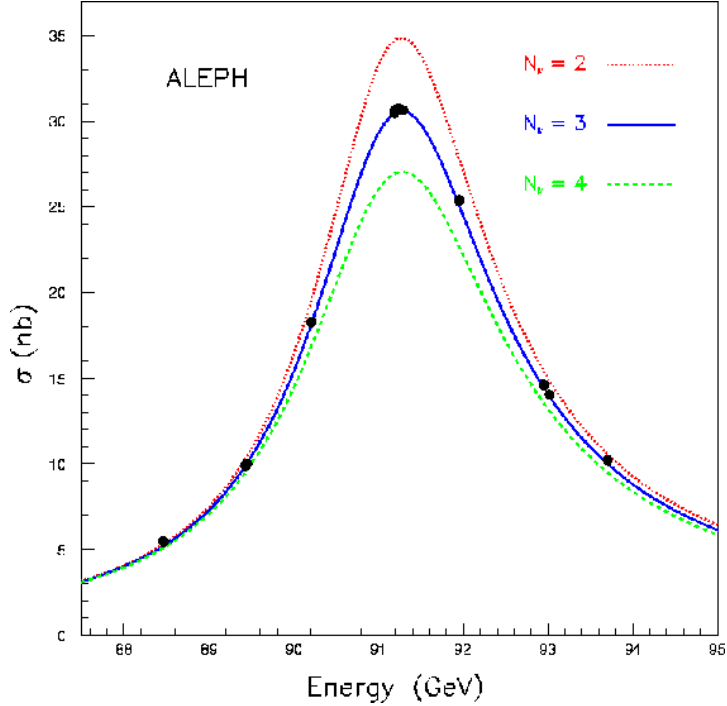}
\caption{\footnotesize
The line-shape of the Z-resonance in $e^{+}e^{-}$ scattering as a
function of the center of mass energy
\protect\cite{Zlineshape:plot,Zlineshape:LEP,Zlineshape:aleph,Zlineshape:opal}.
The theoretical prediction is shown under the assumption that nature
contains $N_\nu$ different neutrinos (Figure taken from
\protect\cite{Zlineshape:plot}).\label{f_zres}}
\end{center}
\end{figure}

Over the last few decades, particle physicists have explored the
fundamental forces down to distance scales of $\approx10^{-18}$\,m. It
was found that the experimental observations are described to very
high accuracy by a theory which is known as the Standard Model of
particle physics. During the 1990s in particular, the predictions of
this theoretical framework have been put to very stringent tests in
accelerator experiments across the world. Perhaps one of the most
impressive examples of its predictive power, the line shape of the
$Z$-resonance in $e^{+}e^{-}$ scattering, is shown in \fig{f_zres}.

The Standard Model describes the interactions of the fundamental
constituents of matter through electromagnetic, weak and strong forces
in terms of three different quantum gauge theories. The success of the
Standard Model is not only a consequence of the mathematical
simplicity of its basic equations, but also because the forces they
describe are relatively weak at the typical energy transfers in
current experiments of about $10-100\,\GeV$.
\footnote{In particle physics it is customary to use ``natural units''
where the speed of light, $c$ and Planck's constant, $\hbar$ are set to one
and energies as well as masses are given in $\GeV$. As an orientation note
that 
$m_{\rm proton}\approx1\,\GeV$, where $1\,\GeV=1.602\cdot10^{-7}\,
{\rm{J}}$.}
The strengths of the interactions are characterized by so-called
coupling constants. When the forces are weak, the predictions of
the theory can be worked out in terms of an expansion in powers of
these coupling constants, a procedure known as perturbation
theory. For instance, in Quantum Electrodynamics (QED), the quantum
gauge theory describing the interactions between electrons and
photons, the coupling constant is the well-known fine structure
constant $\alpha\approx1/137$. Its smallness guarantees that only a
few terms in the power series are sufficient in order to predict
physical quantities with high precision.

The gauge theory for the strong force is called Quantum Chromodynamics
(QCD), in which quarks and gluons assume the r\^oles of the electrons
and photons of QED. Quarks are the constituents of the more familiar
protons and neutrons. The coupling constant of QCD, $\alpha_s$, then
characterizes the strength of the interaction between quarks and
gluons in a similar manner as the fine structure constant does in
QED. One important property of all coupling ``constants'' in
the Standard Model is that they depend on the energy transfer $\mu$ in
the interaction process. In this sense they are not really constant,
and one usually refers to them as couplings that ``run'' with the
energy scale. For instance, at $\mu\approx100\,\GeV$ the strong
coupling constant has been determined as $\alpha_s=0.12$. Although
this is much larger than the fine structure constant of QED, the
method of perturbation theory still works well. However, if the energy
scale $\mu$ is decreased from $100\,\GeV$ it is found that the value
of $\alpha_s$ increases. In fact, at $\mu\approx1\,\GeV$ it becomes so
large that perturbation theory cannot be relied upon any more. It is
then obvious that particle theorists require a tool which is able to
deal with large values of $\alpha_s$. In other words, what is needed
is a {\it non-perturbative} method to work out the predictions of QCD
in this situation.

As mentioned before, the simple and elegant theory of QCD is
formulated in terms of quarks and gluons. Evidence for their existence
has been accumulated in scattering experiments at high energies. Yet
what is observed in experiments at low energies, say,
$\mu\lesssim1\,\GeV$ are protons, neutrons, $\pi$-mesons and many
other particles, all known as hadrons. In fact, a striking property of
QCD is ``confinement'', which means that quarks and gluons cannot be
produced in experiments. Intuitively this property of QCD can be
understood in terms of the strong growth of forces between quarks as
their separation increases. Thus, the only directly observable
particles are the bound states of quarks and gluons, i.e. the hadrons.

Particle physicists are then faced with the task of connecting the
theoretically rather simple regime of QCD at high energies with the
properties of protons, $\pi$-mesons and other hadrons observed at low
energies. This task is made all the more difficult since analytic
methods such as perturbation theory fail completely to describe the
world of hadrons. This has led to the development of numerical
techniques, namely computer simulations of QCD formulated on a
discrete lattice of space-time points. This method allows for a
non-perturbative treatment of the theory in the low-energy regime. 

The basic reason why the two regimes can be connected is the fact that
QCD contains only the fundamental gauge coupling $\alpha_s$ and the
masses of the quarks as free parameters. All observables, such as the
mass of the proton, can in principle be predicted in terms of these
quantities. The ALPHA Collaboration has embarked on a project to
connect low and high energies in practice, by means of extensive
computer simulations of lattice QCD. Following the original idea of
\cite{alpha:sigma}, new methods have been developed and applied, which
have yielded rather precise results. It is then possible to turn the
tables: starting from the accurate experimental information on the
properties of hadrons, the fundamental parameters of QCD can be
determined numerically.

In particular, the high-energy behaviour of the strong coupling
constant $\alpha_s$ is given
by
\bes
   \alpha_s^{-1}
   \,\stackrel{\mu\rightarrow\infty}{\propto}\,
   \ln\left({\mu\over\Lambda}\right) \,,
\ees
This implies that the energy dependence of $\alpha_s$ is specified
completely in terms of a single parameter, $\Lambda$, measured in
energy units.
It is only natural to take $\Lambda$ as a basic
parameter of QCD instead of the energy-dependent $\alpha_s$. The
ratio of $\Lambda$ to the proton mass is computable in lattice
simulations -- but not in perturbation theory. Thus, one of the main
goals of the ALPHA Collaboration is the precise determination of
$\Lambda$.

\begin{wrapfigure}{r}{5.3cm}
\begin{center}
\unitlength 0.35cm
\begin{picture}(15,12)(0,-3)
\linethickness{0.2mm}
\bla\multiput(3,0)(0,2.5){4}{\line( 1, 0){10}}
\bla\multiput(3,0)(2.5,0){4}{\line( 0, 1){10}}
\linethickness{0.4mm}
\multiput(3,2.5)(2.5,0){4}{\bla\circle*{0.2}}
\multiput(3,5)(2.5,0){4}{\bla\circle*{0.2}}
\multiput(3,7.5)(2.5,0){4}{\bla\circle*{0.2}}
\multiput(3,0)(0,2.5){4}{\yel\circle*{0.4}}
\multiput(13,0)(0,2.5){4}{\yel\circle*{0.4}}
\multiput(3,0)(2.5,0){4}{\red\circle*{0.2}}
\multiput(3,10)(2.5,0){4}{\red\circle*{0.2}}

\put( 8.0,7.5){\gre\line(1,0){2.5}}
\put( 8.0,7.5){\gre\vector(1,0){1.3}}

\put(5.5,2.5){\gre\line( 1, 0){2.5}}
\put(8.0,2.5){\gre\line( 0, 1){2.5}}
\put(8.0,5.0){\gre\line(-1, 0){2.5}}
\put(5.5,5.0){\gre\line( 0,-1){2.5}}

\put(5.5,2.5){\gre\vector( 1, 0){1.3}}
\put(8.0,2.5){\gre\vector( 0, 1){1.3}}
\put(8.0,5.0){\gre\vector(-1, 0){1.3}}
\put(5.5,5.0){\gre\vector( 0,-1){1.3}}

\blu\put(2,0){\vector(0,1){10}}
\blu\put(2,10){\vector(0,-1){10}}
\put(0.8,5){$\blu L$}
\mgt\put(13.5,7.5){\vector(0,-1){2.5}}
\mgt\put(13.5,5.0){\vector(0,1){2.5}}
\put(14,6.25){$\mgt a$}

\blu\put( 3,-1){\vector(1,0){10}}
\blu\put(13,-1){\vector(-1,0){10}}
\put(7.6,-2.0){$\blu L$}
\end{picture}

\vspace{-0.3cm}
\caption{\footnotesize
Two dimensional slice of a 4-dimensional space-time lattice. 
Red (yellow) points at the bottom (left) are identified 
with those at the top (right) by periodic boundary conditions.
\label{f_lattice}}
\vspace{-1.2cm}
\end{center}
\end{wrapfigure}

With QCD being one of the pillars of the Standard Model, it is clear
that the precise knowledge of its parameters such as $\Lambda$ is
important for the ongoing quest for generalizations of the Standard
Model. Such more complete theories are
needed to describe the early stages of the universe and are also
expected to be relevant at energies which will be accessible at future
particle colliders.


\section{Lattice QCD}

In the mathematical formulation of QCD, the basic quantities are quark
and gluon fields, which are functions of the space-time coordinates
$x_\mu$, with $x_0$ identified with time. The classical field
equations, which describe their dynamics, are differential equations
-- generalizations of Maxwell's equations for electromagnetism. To
allow for a numerical treatment, it is then natural to discretize the
differential equations with a discretization length $\mgt a$, termed
the lattice spacing.  This procedure turns differential operators into
finite difference operators and the fields are defined only at the
points of a space-time lattice, illustrated in \fig{f_lattice}.

Quantization is achieved by Feynman's path integral representation. It
involves integrations over all degrees of freedom weighted with the
exponential of the classical action. Let $\Omega$ denote an
observable, represented e.g. by a combination of quark and antiquark
fields. Its expectation value, $\langle\Omega\rangle$, is defined as
\be
   \langle\Omega\rangle = {1\over{Z}}\int{\cal D}[U]{\cal D}
   [\psibar,\psi]\,\Omega\,
   \rme^{-S_{\rm G}[U]-S_{\rm F}[U,\psibar,\psi]},
\ee
where $Z$ is fixed by the condition $\langle\unt\rangle=1$. To further
prepare for an evaluation of the path integral on a computer, the
quark degrees of freedom are integrated out analytically. The
expression for $\langle\Omega\rangle$ then becomes
\be
   \langle\Omega\rangle = {1\over{Z}}\int{\cal D}[U]\,\Omega_{\rm eff}
   \left\{\det(D[U])\right\}^{\nf}
   \,\rme^{-S_{\rm G}[U]}.
\label{e_quarkdet}
\ee
$\Omega_{\rm eff}$ denotes the representation of $\Omega$ in the
effective theory, where only gluon fields remain in the path integral
measure. Equation (\ref{e_quarkdet}) requires some further
explanation:
\begin{itemize}
\item 
   $D[U]$ denotes the Dirac operator, and for simplicity of
   presentation we have considered QCD with $\nf$ quarks of equal mass
   $m$, which enters $D[U]$. For hadron physics at small energies the
   case $\nf=3$ is most relevant. The heavier quarks decouple from the
   dynamics up to small effects;
\item
   The gluon field is represented by link variables $U(x,\mu)$
   connecting the sites~$x$ and~$x+a\hat\mu$ (shown as 
   \mbox{\unitlength 0.15cm
         \begin{picture}(3,0.5)(0,0)
         \linethickness{0.2mm}
         \put( 0.2,0.2){\gre\line(1,0){2.5}}
         \put( 0.2,0.2){\gre\vector(1,0){2.0}}
         \end{picture}
   } in Fig.\,\ref{f_lattice}). The measure ${\cal D}[U]$ is the
   product of integration measures for each link;
\item
   The action $S_{\rm G}[U]$ is a sum of local terms over the lattice,
   coupling only gluon variables within one plaquette (
   \mbox{\unitlength 0.15cm
         \begin{picture}(2.5,2.5)(5.2,3.1)
         \linethickness{0.2mm}
         \put(5.5,2.5){\gre\line( 1, 0){2.5}}
         \put(8.0,2.5){\gre\line( 0, 1){2.5}}
         \put(8.0,5.0){\gre\line(-1, 0){2.5}}
         \put(5.5,5.0){\gre\line( 0,-1){2.5}}
         \put(5.5,2.5){\gre\vector( 1, 0){2.0}}
         \put(8.0,2.5){\gre\vector( 0, 1){2.0}}
         \put(8.0,5.0){\gre\vector(-1, 0){2.0}}
         \put(5.5,5.0){\gre\vector( 0,-1){2.0}}
         \end{picture}
    } in \fig{f_lattice}). By contrast, the effective interaction
   resulting from the integration over the anticommuting quark fields
   is of infinite range: although the finite difference Dirac-operator
   $D[U]$ is local, $\det(D[U])$ couples gluons at arbitrary
   distances. 
\item 
   Using the representation \eq{e_quarkdet}, our expression for
   $\langle\Omega\rangle$ has the form of a thermal average in
   statistical mechanics, and one may use stochastic sampling to
   evaluate the integral if the space-time volume is made finite by
   restricting $0\leq x_{\mu} < L$.
\item 
   As an example, consider $\langle \Phi(x)\Phi^\dagger(y) \rangle$
   where $\Phi(x)$ is a suitable combination of quark fields at a
   point $x$, which has the quantum numbers of some hadron. The
   expectation value is then proportional to the quantum-mechanical
   amplitude for the propagation of the hadron from point $y$ to $x$,
   from which the mass of the hadronic bound state may be obtained.
\end{itemize}
The lattice formulation sketched above
leads to a mathematically well-defined
expression for $\langle\Omega\rangle$. In particular, the typical
infinities which are encountered in field theoretical
expectation values are absent. 

The exact treatment of the determinant, \eq{e_quarkdet}, still
presents a major challenge, even on today's massively parallel
computers. In many applications one has therefore set $\nf=0$. This --
very drastic -- approximation defines the so-called {\it quenched
approximation}. Physically it means that the quantum fluctuations of
quarks are neglected and only those due to the gluons are taken
(exactly) into account. Although it turns out that the quenched
approximation works well at the 10\%-level, the proper treatment for
$\nf>0$ is perhaps the most important issue in current simulations.

Whether or not the quenched approximation is employed, one always has
to address the problem of lattice artefacts. Let $R$ denote a
dimensionless observable, such as a ratio of hadron masses. Then its
expectation value on the lattice differs from the value in the
continuum by corrections of order $a^p$:
\bes
  R^{\rm lat} = R^{\rm cont} +O(a^p), 
\label{eq_o_cont_lat}
\ees
where the power~$p$ depends on the chosen discretization of the QCD
action. The correction term for typical values of~$a$ can be quite
large, and an extrapolation to the continuum limit is then required to
obtain the desired result. If~$p$ is large, the rate of convergence to
the continuum limit is high, and hence the extrapolation is much
better controlled.

Let us now return to the problem of determining the energy dependence
of quantities such as the running coupling $\alpha_s$.  Obviously the
numerical simulation is possible only if the number of points of the
lattice, $(L/a)^4$, is not too large; typical lattice sizes are
$L/a\leq 32$. Therefore, besides $a$ also the effect of the finite
value of $L$ has to be considered. It is known rather well that
$L=2\,\fm$ is sufficient for most quantities, and that
much smaller volumes would lead to unacceptably large corrections to
the desired $L\to\infty$ limit. A physical box size of $L=2\,\fm$
together with $L/a\leq 32$ thus implies
\be
 a\;\grtsim\;0.05\,\fm.
\ee
The existence of such a lower bound on $a$, imposed by practicability
considerations, also means that the energies, $\mu$, that can be
expected to be treated correctly have to satisfy
\be
  \mu \ll a^{-1} \lesssim 4\,\GeV\,.
\ee
In other words, lattice QCD is well suited for the computation of low
energy properties of hadrons, while high energies appear impossible to
reach. The ALPHA Collaboration has developed and applied an approach
to circumvent this problem, which is described below.


\section{Running coupling and quark masses}

We are now going to outline the strategy which allows to connect the
low- and high-energy regimes of QCD in a controlled manner. Results
for the $\Lambda$ parameter and the quark masses will then be
presented. From now on we will drop the subscript ``s'' on the
strong coupling constant $\alpha_s$.

\subsection{{Gedankenexperiment}}

From the above it is evident that the restriction of numerical lattice
QCD to low energies is necessary to avoid finite-size effects (FSE)
when working with a manageable number of lattice sites. For the
computation of an effective coupling, this problem can be circumvented
since one has great freedom in the definition of such a coupling. Even
the strength of finite-size effects serve as a measure of the
interactions of the theory, and thus a suitably chosen FSE may be used
to define~\cite{alpha:sigma} an
\bes \red
 \text{effective coupling }\;\gbar(L) \bla \;\text{with}\;
 \alpha(\mu) = {\gbar^2(L) \over 4\pi}\,,\;\red \mu=1/L\,\bla .
\ees
It depends on (``runs with'') the energy scale $\mu=1/L$.  Such a
coupling can be computed in the regime $\mu \ll a^{-1}$ requiring only
a moderate resolution of the space-time world; $L/a = \rmO(10)$ points
per coordinate are sufficient.

In a numerical calculation, the success of this general idea will
depend on a few properties of the coupling. It must be computable with
good statistical precision in a Monte Carlo (MC) simulation, and with
small discretization errors. Furthermore, one would like to know its
scale dependence analytically for large energies. This is achieved by
determining the perturbative expansion of the so-called
$\beta$-function,
\bes
 \beta(\gbar) = -L {\partial\gbar\over\partial L}
     &\buildrel {\gbar}\rightarrow0\over\sim\,&
    -{\gbar}^3 \left \{ b_0 + b_1 {\gbar}^{2} + b_2 {\gbar}^{4} +
    \ldots\right \},
\label{e_beta_pert}
\ees
to the order indicated above, or even higher.

\begin{wrapfigure}{r}{6.5cm}
\begin{center}
\vspace{-0.8cm}
\unitlength 0.35cm
\begin{picture}(17,17)(0,0)
\linethickness{0.3mm}
\bla\put( 1,1){\vector(1,0){12}}
\put(6.6,0.0){ time $t$}
\bla\put( 1,1){\vector(0,1){12}}
\linethickness{0.4mm}
\blu\put(4.6,2){\line( 0, 1){13}}
\blu\put(14.5,2){\line( 0, 1){13}}

\put(5.0,12){\mgt classically:}
\put(5.5,10.5){\mgt electric field}
\put(5.5,9.0){\mgt $E(x)=$ constant }

\linethickness{0.1mm}

\newcommand{\verti}[1]{\begin{turn}{90}#1 \end{turn}}
\linethickness{0.4mm}
\bla
\put(-1.2,5.5){\verti{\mbox{space}}}        
\put( 4.8,7){\red all quantum fields $\phi$:}
\put( 5.5,5.5){\red  $\phi(x+L\hat{k})=\phi(x)$}
\put( 5.5,4){\red  periodic in space}        
\definecolor{lblue}{gray}{.90}
\put(1.9,2){\verti{\colorbox{lblue}{${\bla A_k(x)=}{\blu C\phantom{'}}$ \bla 
                    $\; \psi(x)=0=\psibar(x)$}}}        
\put(14.1,2){\verti{\colorbox{lblue}{${\bla A_k(x)=}{\blu C'}$ \bla 
                    $\; \psi(x)=0=\psibar(x)$}}}        

\end{picture}

\caption{\footnotesize
The boundary conditions used for the definition
of the finite-size coupling $\gbar(L)$ in QCD. 
\label{f_defgbar}}
\vspace{-0.3cm}
\end{center}
\end{wrapfigure}
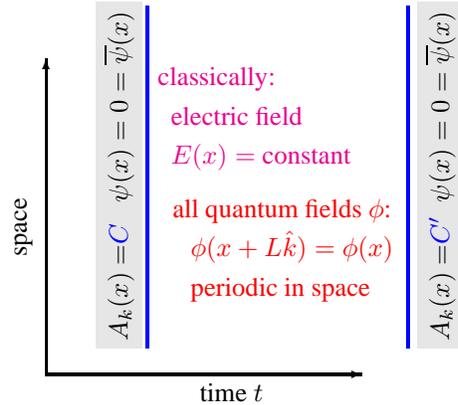

In QCD, a coupling with these properties could indeed be found
\cite{alpha:SU2,alpha:SU3,pert:1loop}.  Its definition starts from QCD
in a box of size $L^4$. Periodic boundary conditions are imposed for
the fields as functions of the three spatial coordinates, and Dirichlet
boundary conditions are set in the time-direction, as shown in
\fig{f_defgbar}. The Dirichlet boundary conditions are homogeneous
except for the spatial components of the gluon gauge potentials,
$A_k$. With this choice of boundary conditions the topology is that of
a 4-dimensional cylinder.

Classically, these boundary conditions lead to a homogeneous ($\vecx$
and $t$ independent) colour-electric field inside the cylinder. The
walls at time $t=0$ and $t=T$ act quite similarly to the plates of an
electric condensor. The strength of the QCD interactions
is conveniently defined in terms of the colour-electric field at the 
condensor plates:
\be
 \gbar^2(L) = {E_{\rm classical} \over \langle E \rangle } \,.
\ee
Here $E$ is a special colour component of the electric field.

 
For weak coupling, i.e. for small $L$, the path integral is dominated
by field configurations which correspond to small fluctuations about
the classical solution. On the other hand, for $L\approx1\,\fm$ they
may deviate significantly from the classical solution, and this can be
realized in a MC-simulation of the path integral.


\subsection{The running coupling}

\begin{wrapfigure}{r}{8.0cm}
\begin{center}
\vspace{-0.5cm}
\input{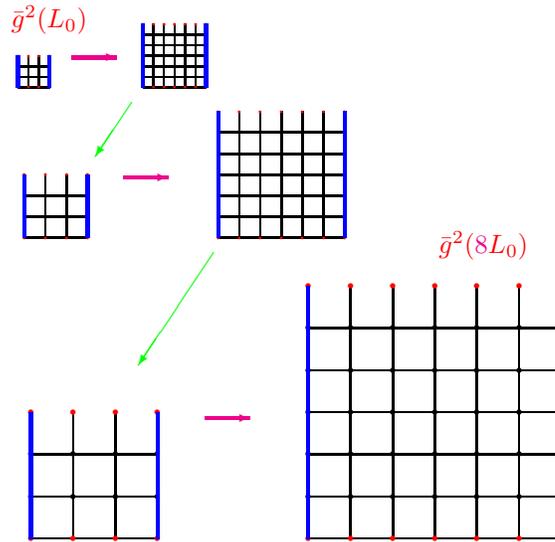}
\caption{\footnotesize
The recursion to compute $\gbar(L)$ with
$L=L_0 \mgt \to \bla 2 L_0 \mgt \to \bla 4  L_0 \mgt \to \bla 8  L_0$.
It is iterated further until one arrives at  $2^n L_0$,
in practice $n \sim 8$. One always alternates between
{\mgt keeping $a$ fixed and increasing $L \to 2L$} and
{\gre keeping $L$ fixed and increasing $a$}.
\label{f_recursion}}
\vspace{0.7cm}
\end{center}
\end{wrapfigure}

The energy dependence of the coupling $\gbar^2(L)$ can be determined
recursively through a number of compute-effective steps.
\Fig{f_recursion} illustrates the implementation of the recursion
\bes
  \gbar^2(2L) = \sigma(\gbar^2(L))\,.
\label{e_recursion}
\ees
%
The function $\sigma$ describes the change in the coupling when the
physical box size is doubled. Since the relevant energy scale for the
running of the coupling ($\mu=1/L$) is separated from the lattice
spacing~$a$, one may compute $\sigma$ recursively over several orders
of magnitude in $\mu$, whilst keeping the number of lattice sites at a
manageable level, i.e. $L/a=\rmO(10)$.

\begin{wrapfigure}{r}{7.0cm}
\begin{center}
\includegraphics[width=7.0cm]{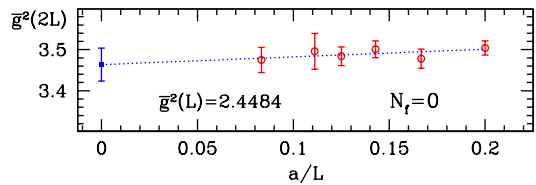}
\vspace{-0.4cm}
\caption{\footnotesize
The extrapolation of the function $\gbar^2(2L) = \sigma(\gbar^2(L))$
from the data at finite lattice spacing $a$ to the continuum (blue square).
\label{f_Sigma}}
\vspace{-0.5cm}
\end{center}
\end{wrapfigure}

The results of one horizontal step in Fig.~\ref{f_recursion},
performed on the APE-computers at NIC/Zeuthen, is displayed in
\fig{f_Sigma}.  They are shown as a function of the resolution
$a/L$. Because the coupling and the details of the discretization were
chosen with great care, the dependence on the resolution is tiny and
the lattice numbers can be extrapolated to the {\em continuum} $a/L\to
0$. To arrive at the continuum limit, each of the horizontal steps in
the figure is indeed repeated several times with different
resolutions!

\begin{figure}[t]
\hspace{-1cm}
{\includegraphics[width=8.0cm]{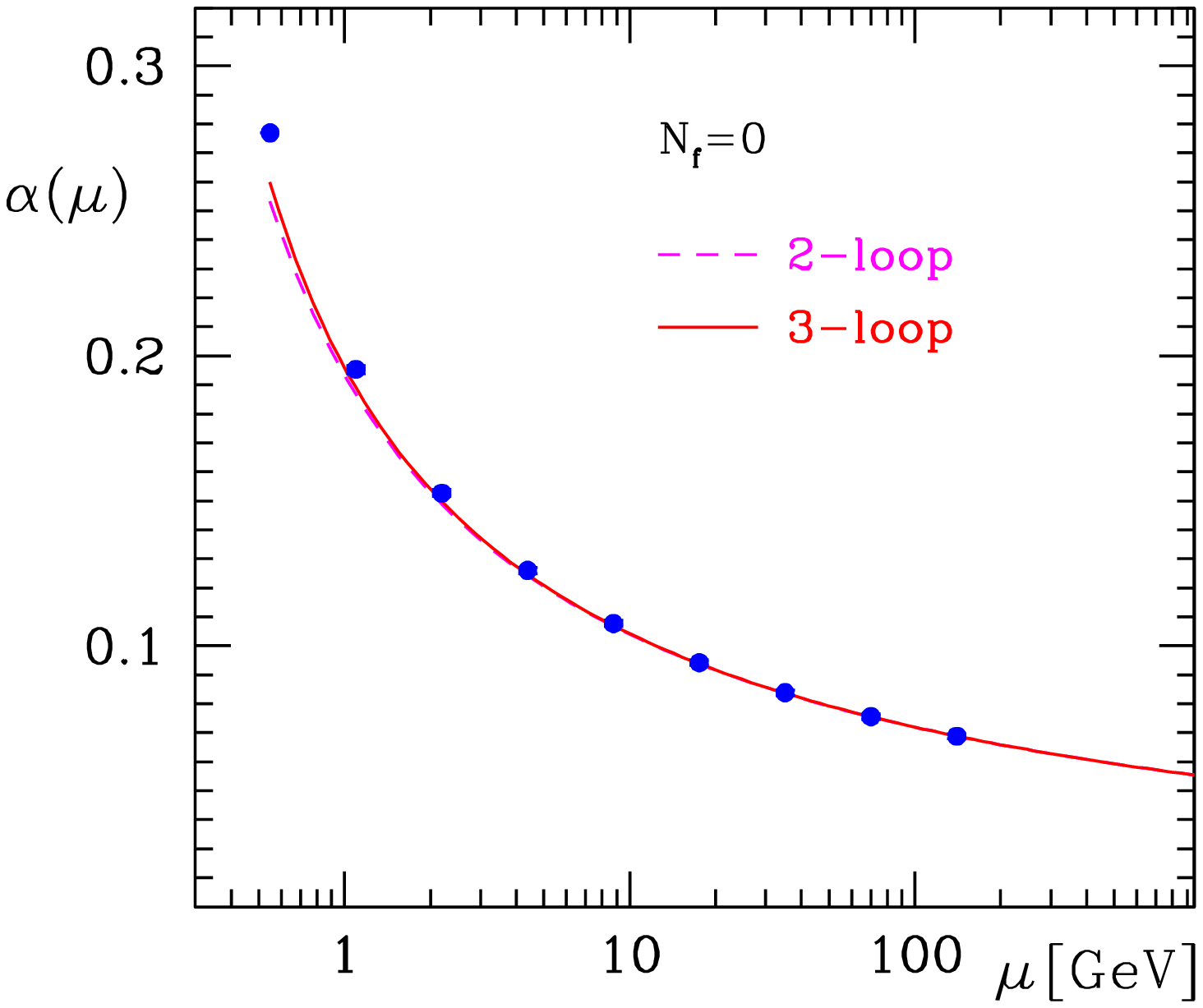}}
\vspace{-8.0cm}
\hspace{-2cm}
{\includegraphics[width=8.0cm]{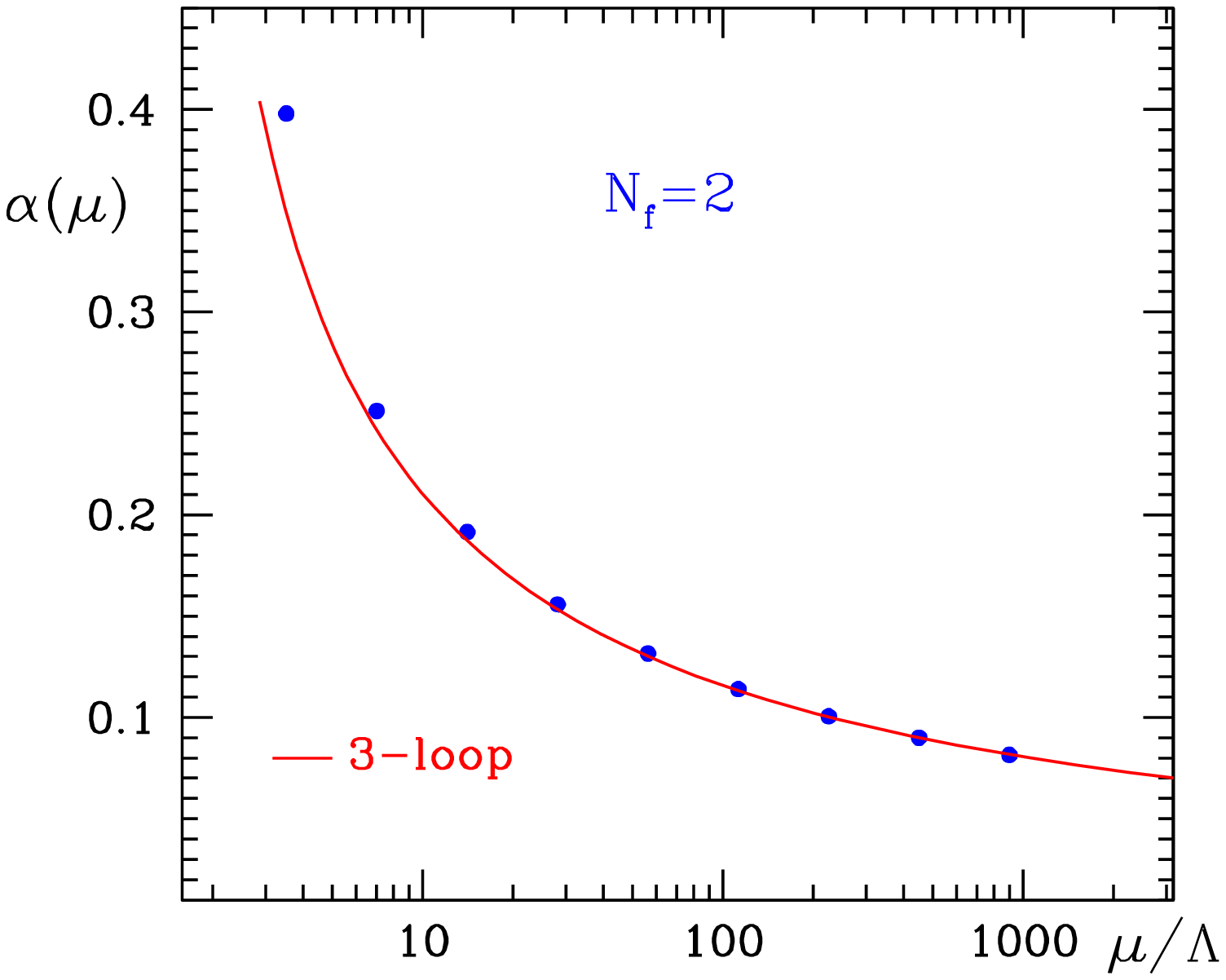}}
\vspace{6.0cm}
\caption{\footnotesize
The energy dependence of the coupling as computed from lattice
QCD (points) compared to the perturbative expression
\protect\eq{e_beta_pert} with the $b_2$-term (full line) and without.
\label{f_alpha}}
\end{figure}

In this way, the continuum function $\sigma(u)$ was obtained for a
range of $u=\gbar^2(L)$. Starting from a minimal value of $u$, the
recursive application of \eq{e_recursion} yields the points in
\fig{f_alpha}. They may be compared to the integration of the
differential equation (\ref{e_beta_pert}) towards low values of $\mu$,
starting at the lowest value of $\alpha$. One then makes the important
observation, that \eq{e_beta_pert}, truncated at the indicated order,
is quantitatively verified in the region of low enough $\alpha$. It
may hence also be used to compute the $\Lambda$-parameter from
($\gbar=\gbar(L=1/\mu)$)
\bes 
\Lambda &=& \mu\,(b_0\gbar^2)^{-b_1/2b_0^2} \rme^{-1/2b_0\gbar^2}
  \times \exp\left\{-\int_0^{\gbar}\rmd g
  \left[{1\over\beta(g)}+{1\over b_0g^3}-{b_1\over b_0^2 g}\right]\right\},
\label{f_Lambda_def} 
\ees
with negligible errors due to higher-order terms that are not
included.

The attentive reader will have noticed that the computation explained
so far, ``only'' determines the dependence of $\alpha$ on the
combination $\mu/\Lambda$, while in the figure we show it as a
function of $\mu$ in physical units for $\nf=0$, i.e. in the quenched
approximation. The missing link is to connect the lowest energy
$\mu=1/L_{\rm max}$ contained in the figure, to a low energy --
experimentally accessible -- property of a hadron. The connection with
the decay rate of the K-meson~\cite{mbar:pap3}, which we do not
describe here, yields the physical units shown in the figure as well
as the final result
\bes
 \Lambda = 238(19)\,\MeV \text{ for } \nf=0\,.
\ees  
Currently, the ALPHA Collaboration is extending these computations to
the numerically very demanding case of $\nf=2$. First promising
results, which are still awaiting a final check for the absence of
discretization errors are shown on the r.h.s.  of \fig{f_alpha}. The
step to connect $L_{\rm max}$ to a low-energy observable has yet to be
performed, and hence the results are plotted as a function of
$\mu/\Lambda$. It is worth pointing out that, in order to ensure
efficiency and correctness of these $\nf>0$ simulations, the
development and testing of MC-algorithms is very
important\,\cite{alpha:bench,algo:MH,algo:GHMC}.

\subsection{Quark masses}

The masses of the quarks are of a very different nature from the mass
of the electron. The fundamental difference is the property of
confinement mentioned in section 1. It means that a quark can not be
prepared in isolation to perform an experimental measurement of its
mass. Consequently, quark masses have to be understood as parameters
of the theory, and their proper definition very much resembles that of
an effective coupling. In particular, they have an energy dependence
similar to the one of $\alpha$. When defined in a natural way
\cite{mbar:pap1}, the energy dependence is the same for all quark
flavours; in other words, ratios of quark masses do not depend on the
energy. The overall energy dependence has been determined with the
same strategy as the one described for $\alpha$ and in fact with
similar precision (see Fig.\,2 in \cite{mbar:pap1}).

\begin{figure}
\begin{center}
{\includegraphics[width=8.0cm]{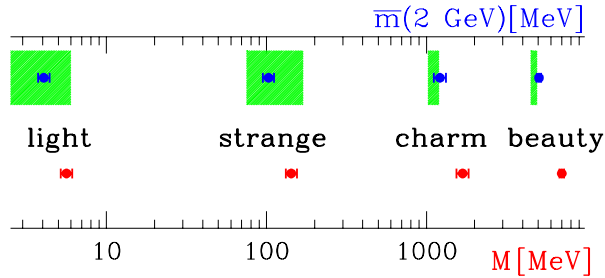}}
\caption{
Quark masses determined by the ALPHA collaboration in the $\Nf=0$
approximation
\protect\cite{mbar:pert,mbar:pap1,mbar:pap3,mcharm:lat01,mb:lat01}.
The label ``light'' denotes the average mass of the up and the down
quark mass. Green shaded areas show the error band quoted by the
particle data group \protect\cite{PDG00}.
\label{f_quarkmasses}}
\end{center}
\end{figure}

The numerical values of the quark masses (at a particular low energy
scale) are conveniently extracted by relating them to the masses of
quark-antiquark bound states with pseudoscalar quantum numbers:
$\pi$-, K-, D-, B-mesons. Results of the ALPHA Collaboration obtained
in the $\nf=0$ approximation are shown in \fig{f_quarkmasses} in two
different (common) conventions.~\footnote{The ratio of strange to
light quark masses actually relies on \cite{leutwyler:1996}. In the
special case of the b-quark, whose mass is large compared to the QCD
scale, $\Lambda$, an expansion in terms of $1/m_{\rm b}$ has been used
and only the lowest order term was considered~\cite{mb:lat01}.} The
upper row shows the running masses $\mbar$ in the so-called $\msbar$
scheme at a renormalization energy $\mu=2\,\GeV$, while the lower one
shows the so-called renormalization group invariant quark masses. The
latter are related to the running quark masses via 
\bes
  M =\lim_{\mu\to\infty} \mbar(\mu)
  \left[2b_0\left(\gbar(\mu)\right)^2\right]^{-d_0/2b_0}\,,
\ees
where $d_0$ characterizes the asymptotic behavior of $\mbar(\mu)$ for
large energy. It is computable in perturbation theory and has a value
of $d_0=8/(4\pi)^2$.


\section{Improvement}

Equation (\ref{eq_o_cont_lat}) shows that lattice and continuum
quantities differ by discretization errors. In Wilson's formulation of
lattice QCD these discretization effects are of order~$a$, which can
be quite large. People have therefore tried to find an improved
version of the Wilson action, so that the deviation of observables
computed on the lattice to their continuum values is of O($a^2$), or
even higher.

In order to illustrate the problem, let us consider the solution of
differential equations by means of numerical integration.
Given the initial value problem
\bes
   {{\rmd y}\over{\rmd x}} = f(x,y),\quad y(x_0) = y_0,
\ees
the well-known Euler method provides a formula for its numerical
solution, i.e.
\bes
   y_{j+1} = y_j +a\,f(x_j,y_j),\quad j=0,1,2,\ldots.
\label{eq_soldeq}
\ees
Here the quantity~$a$ denotes the finite step size used to approximate
the derivative ${\rmd y}/{\rmd x}$ in terms of finite differences. In
order to compute the solution at point $x=x_0+na$ for a fixed value
of~$a$ one has to perform~$n$ iteration steps. The central question
then is by how much this solution deviates from the exact one. For the
simple Euler method one can show that this so-called truncation error
is proportional to~$a^2$. Obviously the accuracy of the solution will
be larger for small~$a$, but then the number of iterations, $n$, may
become very large. It is much more efficient to use an improved
solution scheme, such as the Runge-Kutta method. Here the function $f$
in \eq{eq_soldeq} is replaced by a more complicated expression, which
is chosen in such a way that the truncation error is proportional
to~$a^5$.

The analogy of this example with lattice QCD is obvious: the step size
is now the lattice spacing, and the lattice action must be chosen such
that the leading discretization error is of higher order in~$a$.
However, in a quantum field theory like QCD the search for an improved
action is not straightforward: although one can quite easily write
down an action which formally has O($a^2$) artefacts, the interactions
between quarks and gluons in the discretized theory can again
introduce O($a$) errors at the quantum level.

To be more explicit, let $S_{\rm W}$ denote the Wilson action for
lattice QCD. Sheikholeslami and Wohlert \cite{impr:SW} have shown that
it is sufficient to add just one more interaction term, corresponding
to an action $\delta{S}$, to $S_{\rm W}$ in order to cancel
its leading discretization error proportional to~$a$. To ensure that quantum
effects do not introduce O($a$) effects ``via the back door'', the
term $\delta{S}$ has to be multiplied by a coefficient $\csw$. This
improvement coefficient must be suitably tuned in order to
achieve the complete cancellation of O($a$) artefacts at the quantum
level for observables such as hadron masses. The O($a$) improved
version of Wilson's action then reads
\bes
   S_{\rm SW} = S_{\rm W}+\csw\delta{S}.
\ees
The ALPHA Collaboration has developed and applied a method to
determine $\csw$ in computer simulations
\cite{impr:pap1,impr:pap2,impr:pap3,impr:csw_nf2}. The method amounts
to computing the expectation value of a pure lattice artefact as a
function of $\csw$. The value of $\csw$ at the point where the
artefact vanishes then defines the improved action for the
corresponding value of the lattice spacing. Thus, non-perturbative
methods are not only used to compute physical observables, but also to
improve the reliability of the numerical treatment of the theory as
such. The improved action is only slightly more complicated to
implement in practical simulations. The increased effort is easily
offset by the faster convergence to the continuum limit. As a result,
the action with $\csw$ as determined in \cite{impr:pap3,impr:csw_nf2}
has by now become a standard for precision lattice QCD
computations. In particular this improvement was essential in
obtaining most of the results quoted in the previous section.

Of course, one also has to verify that observables computed with the
improved action indeed approach the continuum limit with a rate
proportional to~$a^2$. Tests of this kind have been performed
successfully (see, e.g. ref. \cite{impr:jochen}), and other examples
for the effectiveness of the improvement programme for the evaluation
of many hadronic observables can be found in~\cite{mbar:pap3}. 


\section{Machines and resources}

Although conceptual advances like O($a$) improvement are of great
importance, progress in lattice QCD is also dependent on the
availability of sufficient computer resources.

In the 1980s most simulations were performed on vector supercomputers
like the Cray X-MP. Since then the demand for increased performance
has led to the development of massively parallel machines, which are
now widely used. Lattice QCD is a problem which lends itself easily to
parallelization: the total volume can be divided into many
sublattices, which are distributed over a grid of processors. The
latter can perform the same task on independent data.  Furthermore,
the communication pattern for lattice QCD is simple, since most
algorithms only require nearest-neighbour communications.

The ALPHA Collaboration has mostly used parallel computers from the
APE family of
machines~\cite{APE100_HW,APE100_SW,APEmille_stat,APEmille_lat01}. The
latest generation, APEmille, has been developed jointly by INFN and
DESY. The smallest entity in the APEmille processor grid consists of a
cube of $2\times2\times2$ processors. These cubes are then connected
to form larger grids. Despite its very conservative clock speed of
66\,MHz, each processor achieves a peak performance of
528\,MFlops,
\footnote{The unit 1\,MFlops denotes one million floating
point operations per second.}
thanks to the optimization for complex arithmetic: the operation
$a\times{b}+c$, which requires~8 floating point operations for the
three 32-bit complex numbers $a,\,b$ and~$c$, is performed in one
clock cycle. The total peak speed of the current installation of
APEmille machines at DESY-Zeuthen amounts to more than
500\,GFlops (single precision). 

The programming language for APEmille is TAO, which has a FORTRAN-like
syntax, but also includes special features designed to facilitate
parallelization and coding, and allows to achieve a high proportion of
the peak speed. One such feature is the easy access to any desired
number of the 512(!) registers. The typical efficiency of
ALPHA-programs on APEmille is about 30\% of the peak speed. A similar
figure has been achieved by other lattice QCD collaborations on
machines like the Cray T3E, but at the price of having to code the
core routines in assembler language~\cite{perf:UKQCD}. An interesting
figure-of-merit is the price/performance ratio. For APEmille it is
$(8-10)$ Euro per sustained MFlops and there are efforts under way to
reduce this number much further. In addition it was also demonstrated
that PC clusters with Myrinet network can achieve about $3-4$
Euro/MFlops sustained \cite{ML_lat01}. These developments let us await
the future with optimism.

\section{Future} 
Moving towards the realistic case of $\nf=3$ should soon be
possible. Once this has been achieved, the most precise determinations
of the fundamental parameters of QCD may come from the low energy
hadron spectrum combined with lattice QCD to evaluate the theory. The
methods developed in this project are expected to play an important
r\^ole in this program.

Furthermore these methods and related
ones~\cite{tmQCD:pap1,tmQCD:pap2,tmQCD:BK1} will improve the
reliability of the determination of properly normalized weak decay
(and mixing) amplitudes of hadrons~\cite{lat01:marti}. Their knowledge
is again of vital importance in tests of the Standard model and the
search for an even more fundamental theory.

\vspace{1.0cm}

The results presented in this article were obtained on the APE100 and
APEmille installations at Zeuthen. The total CPU time required was
about $10^7$ processor-hours on APE100 and $0.2\cdot10^7$
processor-hours on APEmille, respectively.



\end{document}